\documentclass{article}

\usepackage{PRIMEarxiv}

\usepackage[utf8]{inputenc} 
\usepackage[T1]{fontenc}    
\usepackage{hyperref}       
\usepackage{url}            
\usepackage{booktabs}       
\usepackage{amsfonts}       
\usepackage{nicefrac}       
\usepackage{microtype}      
\usepackage{lipsum}
\usepackage{fancyhdr}       
\usepackage{graphicx}       
\usepackage{multirow}

\usepackage{amssymb}
\usepackage{amsmath}
\usepackage{stfloats}
\usepackage{xcolor}
\usepackage{algorithm}  
\usepackage{algorithmic}
\usepackage[normalem]{ulem}
\usepackage{lscape}
\usepackage{geometry}
\usepackage{caption} 
\usepackage{longtable}
\usepackage{setspace} 

\usepackage{pifont}
\newcommand{\cmark}{\ding{51}}%
\newcommand{\xmark}{\ding{55}}%

\usepackage[justification=centering]{caption}

 \usepackage{lineno}

\graphicspath{{media/}}     

\title{Neural-based Cross-modal Search and Retrieval of Artwork
}

  \author{
  Yan Gong, Georgina Cosma, Axel Finke \\
  Loughborough University \\
  Loughborough\\
  \texttt{\{y.gong2, g.cosma, a.finke\}@lboro.ac.uk}
  }

\begin{document}
\maketitle

\begin{abstract}
Creating an intelligent search and retrieval system for artwork images, particularly paintings, is crucial for documenting cultural heritage, fostering wider public engagement, and advancing artistic analysis and interpretation. Visual-Semantic Embedding (VSE) networks are deep learning models used for information retrieval, which learn joint representations of textual and visual data, enabling 1) cross-modal search and retrieval tasks, such as image-to-text and text-to-image retrieval; and 2) relation-focused retrieval to capture entity relationships and provide more contextually relevant search results. Although VSE networks have played a significant role in cross-modal information retrieval, their application to painting datasets, such as ArtUK, remains unexplored. This paper introduces BoonArt, a VSE-based cross-modal search engine that allows users to search for images using textual queries, and to obtain textual descriptions along with the corresponding images when using image queries. The performance of BoonArt was evaluated using the ArtUK dataset. Experimental evaluations revealed that BoonArt achieved 97\,\% Recall@10 for image-to-text retrieval, and 97.4\,\% Recall@10 for text-to-image Retrieval. By bridging the gap between textual and visual modalities, BoonArt provides a much-improved search performance compared to traditional search engines, such as the one provided by the ArtUK website. BoonArt can be utilised to work with other artwork datasets.

\end{abstract}


\section{Introduction}
Creating a retrieval system for artwork images, particularly paintings, is of paramount importance in documenting cultural heritage, facilitating wider public engagement, and fostering advancements in art analysis and interpretation \cite{castellano2021visual, li2022intuitively}. Existing research for retrieving painting images primarily focuses on using neural networks to classify objects within painting images and match their categories with user queries for retrieval purposes \cite{crowley2014state, crowley2016art, seidenari2017deep}. However, these studies have limitations in fully comprehending the complex semantic information expressed in user queries. While research specific to information retrieval can uncover the underlying meaning of user queries, text-based information retrieval \cite{lowe1999object, zhang2013image, li2011text} lacks the ability to understand the visual content of images, relying solely on textual information such as image tags. On the other hand, cross-modal information retrieval is capable of extracting and comprehending high-level visual semantics in conjunction with textual information, enabling users to obtain more relevant and accurate results \cite{kaur2021comparative}, which is appropriate for applying to retrieval of painting images. Visual-Semantic Embedding (VSE) networks represent state-of-the-art techniques in cross-modal information retrieval. These networks aim to embed image-description pairs into a shared latent space, enabling the computation of similarity scores for image-to-text and text-to-image retrieval tasks \cite{zhang2020context, li2020unicoder}. VSE networks have demonstrated their effectiveness with real-world images in widely used benchmark datasets such as Flickr30K \cite{young2014image}, MS-COCO \cite{lin2014microsoft}, and RefCOCOg \cite{mao2016generation}. Despite their successful application in these datasets, their potential for implementing retrieval of painting images, especially in the context of being integrated into a cross-modal search engine, remains unexplored, representing an untapped area of research.

\begin{figure}[htbp] 
\centering 
\includegraphics[width=0.42\linewidth]{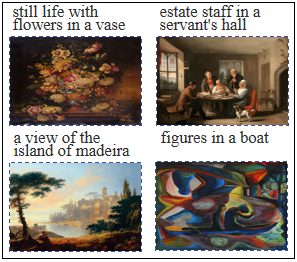} 
\caption{Examples of painting images with corresponding textual description.} 
\label{ArtUKdata} 
\end{figure} 

In recent years, there have been significant advancements in the field of VSE. 
Faghri et al.\ \cite{faghri2018vse++} introduced an architecture that embeds image region features extracted by the faster R-CNN \cite{anderson2018bottom} and their descriptions into a shared latent space using a fully connected neural network and a Gated Recurrent Units (GRU) network \cite{cho2014learning}, respectively. 
Lee et al.\ \cite{lee2018stacked} augmented VSE networks by employing the attention mechanism to enhance the alignment of image regions with their corresponding words.
Li et al.\ \cite{li2019visual} proposed the visual semantic reasoning network, which leverages the Graph Convolution Network (GCN) \cite{kipf2017semi} to extract high-level visual semantics. 
Chen et al.\ \cite{chen2021learning} unveiled a network employing a generalised pooling operator to formulate an optimal strategy for integrating image and description representations. 
Chen et al.\ \cite{chen2020uniter} introduced a pre-trained network built upon the transformer model \cite{devlin2019bert} that has been pre-trained on four large datasets \cite{lin2014microsoft, krishna2017visual, sharma2018conceptual, ordonez2011im2text}. 
Recently, Radford et al.\ \cite{radford2021learning} proposed a Contrastive Language-Image Pre-training network (CLIP), which leverages 400 million image-description pairs to enable the efficient learning of visual concepts via natural language supervision. 
To overcome the limitation of the pre-trained Vision Transformers (ViTs) for relation-focused cross-modal information retrieval, Gong et al.\ \cite{gong2023vitr} proposed a ViT-Relation-focus network (VITR), which employs a local encoder to reason about relations within image regions. 
This paper primarily focuses on implementing a VSE-based cross-modal search engine designed for retrieving painting images. Figure~\ref{ArtUKdata} showcases a variety of painting images, encompassing different styles including realistic, impressionistic, abstract, and still life paintings. The proposed search engine, named BoonArt, incorporates a VSE network, VITR, enhancing its capability in image-to-text and text-to-image retrieval tasks within the painting images, with a particular emphasis on relation-focused cross-modal information retrieval. Specifically, the contributions are as follows.
\begin{itemize} 
    \item BoonArt search engine has the capability to perform image-to-text and text-to-image retrieval for painting images; that BoonArt allows users to search for painting images using textual queries and retrieve textual descriptions along with their corresponding images using image queries; and that BoonArt benefits from a state-of-the-art VSE network, VITR, which can extract and understand high-level visual semantics to improve retrieval performance and enhance the user experience.
    
    \item BoonArt's performance is evaluated through experiments using the ArtUK dataset. The results demonstrate that BoonArt outperformes the ArtUK search system (from the ArtUK website https://artuk.org) for text-to-image tasks. In particular, BoonArt can use image queries for retrieval. In contrast, the ArtUK search system lacks this capability.
\end{itemize} 

\section{Methodology}

BoonArt excels at retrieving painting images from the ArtUK dataset \cite{elliot2021artuk}.  Users can enable text-to-image requests to search for relevant images based on their textual queries. Additionally, users can perform image-to-text requests to search for descriptions and their corresponding images using image queries.
The BoonArt engine is composed of three main components: the front-end, back-end, and a database (images, textual descriptions, and representations, i.e., embeddings). Each component functions as follows. 

\subsection{Front-end}
\begin{figure*}[htbp] 
\centering 
\includegraphics[width=1\linewidth]{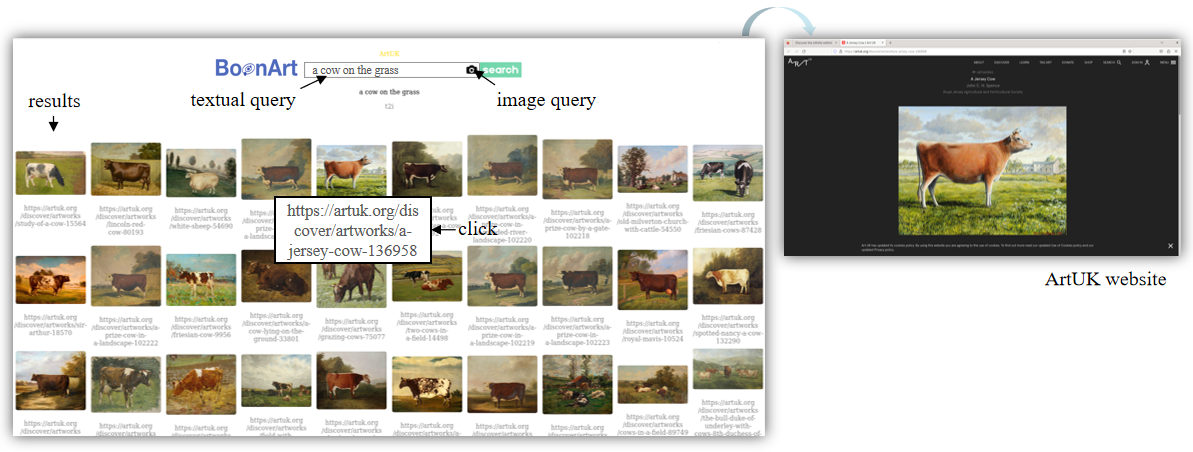} 
\caption{The front-end of BoonArt enables users to input textual queries for text-to-image retrieval or upload image queries for image-to-text retrieval. The search results are displayed, allowing users to explore painting images, with additional details accessible through clicking links to the ArtUK website.} 
\label{FlowArtUK} 
\end{figure*}

As depicted in Figure~\ref{FlowArtUK}, in the front-end, users are presented with two options for querying: they can either input a textual query in the provided text box to enable the text-to-image request, or they can upload an image query using the designated upload button to enable the image-to-text request. 
After entering their query, users can initiate the search by clicking the search button. 
The retrieval results are then displayed on the interface, allowing users to view and explore the painting images. To provide additional information about the retrieved paintings, users can click on the links associated with the images, which will open a new page on the source website of each painting. 
Users can input their queries with high-level semantics, such as a focus on relations, and BoonArt will provide accurate search results based on the semantics of their queries.

\subsection{Back-end}

\begin{figure}[htbp] 
\centering 
\includegraphics[width=0.5\linewidth]{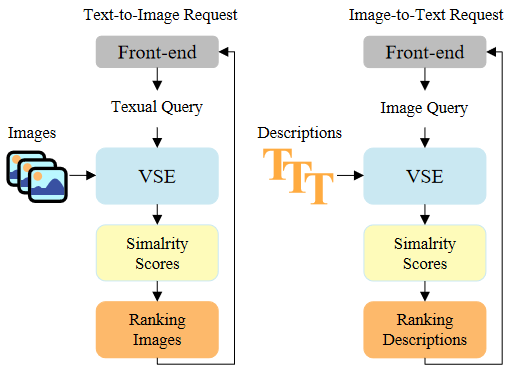} 
\caption{The back-end of BoonArt utilises a VSE network for text-to-image and image-to-text requests, comparing queries with the dataset's images or descriptions and generating ranked results displayed by the front-end.} 
\label{BackEndArtUK} 
\end{figure}

The core of BoonArt's back-end is a VSE network, and the process of back-end is shown as Figure~\ref{BackEndArtUK}. 
For the text-to-image request, the VSE network compares the textual query with all the images in the dataset. It generates similarity scores for each image and ranks them accordingly. The front-end then displays the ranked images. 
Similarly, for the image-to-text request, the VSE network compares the image query with all the descriptions in the dataset. It calculates similarity scores for each description and ranks them accordingly. The front-end displays the images corresponding to the ranked descriptions.

BoonArt employs the state-of-the-art VSE network VITR \cite{gong2023vitr}. 
VITR consists of a text encoder for encoding descriptions as global and local representations, a ViT encoder for encoding images as global representations, a local encoder for encoding images as local representations for relational reasoning, and a fusion module that fuses the representations from the encoders to output the similarity score between the image and the description. 
VITR takes the text and ViT encoders from CLIP to obtain pre-trained knowledge from an extensive dataset of 400 million image-description pairs. Additionally, VITR has been fine-tuned on the RefCOCOg dataset, which enhances the network's ability to learn the reasoning relations of image regions to improve performance in relation-focused cross-modal information retrieval tasks. 

\subsection{Database}
The database is built upon the ArtUK dataset \cite{elliot2021artuk}, which consists of 6\,783 pairs of painting images and their descriptions, sourced from the ArtUK website (https://artuk.org).   
To optimise retrieval time, the representations for images and descriptions needed by the VSE network have been pre-encoded and stored. The following files were created and are available in the database: 
1) ArtUKimGloRp.npy (13.0\,MB) stores the global representations of images; 
2) ArtUKimLocRp.npy (1.7\,GB) stores the local representations of images; 
3) ArtUKdeGloRp.npy (13.0\,MB) stores the global representations of descriptions; 
and 4) ArtUKdeLocRp.npy (999.8\,MB) file stores the local representations of descriptions. 
By directly accessing the saved representation values from these files, the back-end eliminates the need for encoding images and descriptions during the retrieval process, resulting in faster retrieval.

\section{Experiments}
\subsection{Implementation Details}
BoonArt can function with a minimum requirement of a single NVIDIA RTX 3080 graphics card. The integrated VSE network of BoonArt is VITR, which leverages the encoder of `ViT-L/14' from CLIP. To maintain generalizability in real-world image scenarios, BoonArt's VITR employs zero-shot learning on the ArtUK dataset. 

\begin{figure*}[htbp] 
\centering 
\includegraphics[width=0.95\linewidth]{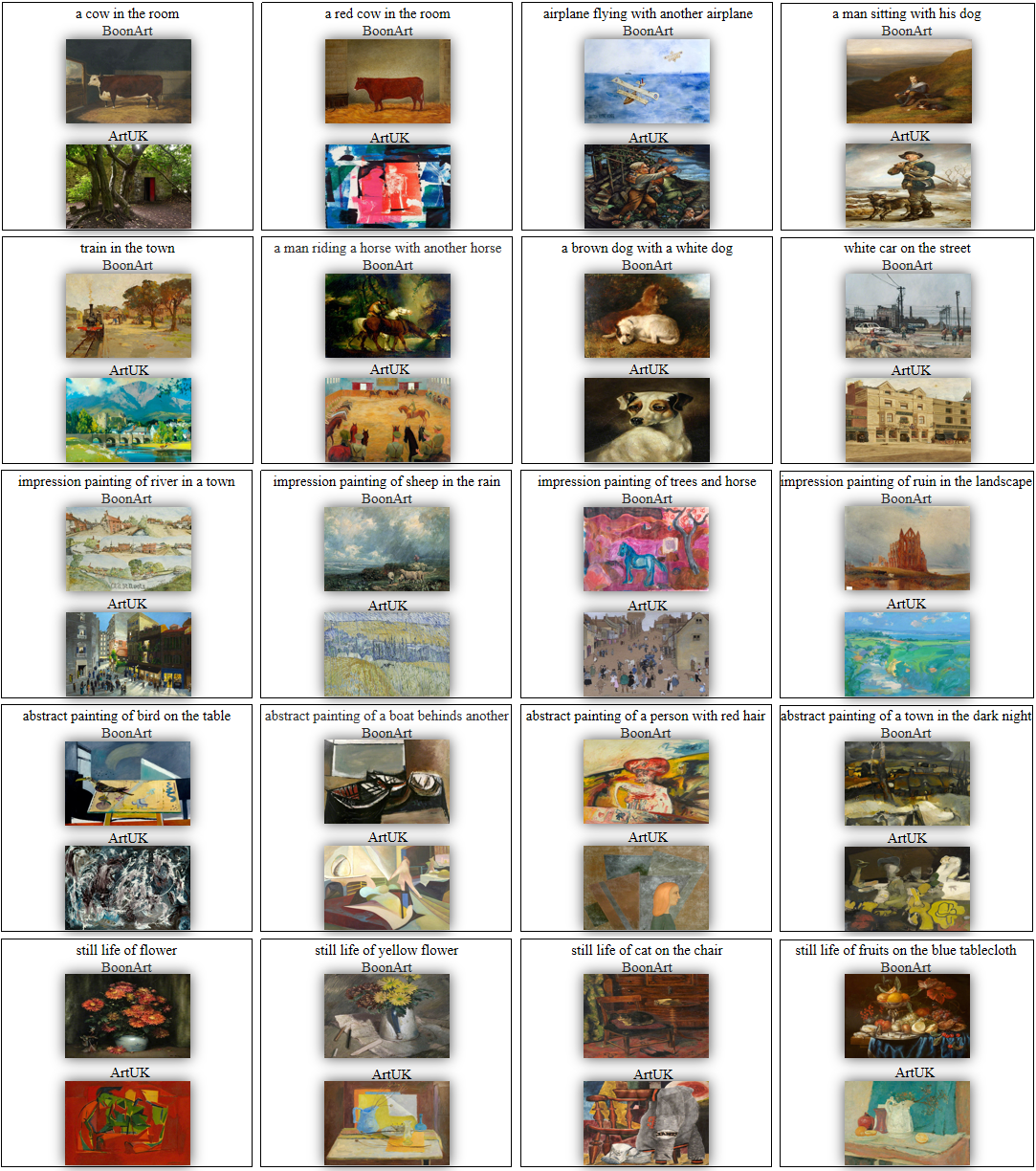} 
\caption{Comparison between BoonArt and the ArtUK search system for text-to-image retrieval. The figure shows the top-ranked retrieved images for the queries.} 
\label{FigureComparisonBoonArtArtUK} 
\end{figure*} 

\begin{figure*}[htbp] 
\centering 
\includegraphics[width=0.95\linewidth]{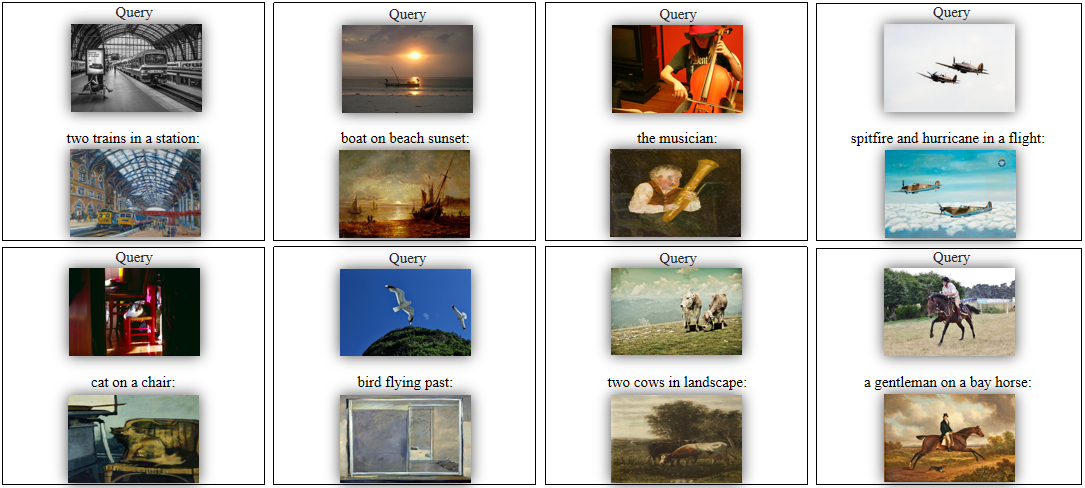} 
\caption{Retrieval results of BoonArt for image-to-text retrieval. The figure shows the top-ranked retrieved descriptions and the images corresponding to the queries.} 
\label{FigureComparisonBoonArtArtUKi2t} 
\end{figure*} 

\subsection{A Comparison of BoonArt and the ArtUK Search System for Text-to-Image Retrieval}

\begin{table}[ht]
\caption{Performance of BoonArt with the ArtUKA search system on a set of textual queries.}
\label{TableComparisonBoonArtArtUK}
\centering
\begin{tabular}{lcc}
\hline
Query & \begin{tabular}[c]{@{}c@{}}BoonArt\end{tabular} & \begin{tabular}[c]{@{}c@{}}ArtUK\end{tabular} \\ \hline
1. a cow in the room & \cmark & \xmark \\
2. a red cow in the room &  \cmark & \xmark  \\
3. airplane flying with another airplane &  \cmark & \xmark  \\
4. a man sitting with his dog &  \cmark & \xmark  \\
5. train in the town &  \cmark & \xmark  \\
6. a man riding a horse with another horse &  \cmark & \xmark  \\
7. a brown dog with a white dog &  \cmark & \xmark  \\
8. white car on the street &  \cmark & \xmark  \\
9. impression painting of river in a town &  \cmark & \xmark  \\
10. impression painting of sheep in the rain &  \cmark & \xmark  \\
11. impression painting of trees and horse &  \cmark & \xmark  \\
12. impression painting of ruin in the landscape &  \cmark & \xmark  \\
13. abstract painting of bird on the table &  \cmark & \xmark  \\
14. abstract painting of a boat behinds another &  \cmark & \xmark  \\
15. abstract painting of a person with red hair &  \cmark & \xmark  \\
16. abstract painting of a town in the dark night &  \cmark & \xmark  \\
17. still life of flower &  \cmark & \xmark  \\
18. still life of yellow flower &  \cmark & \xmark  \\
19. still life of cat on the chair &  \cmark & \xmark  \\
20. still life of fruits on the blue tablecloth &  \cmark & \xmark  \\ \hline
\end{tabular}
\end{table}

To compare the performance between BoonArt and the ArtUK search system (from the ArtUK website) for text-to-image retrieval, a set of 20 textual queries was generated, as shown in Table~\ref{TableComparisonBoonArtArtUK}. These queries encompass high-level semantics, including relations. Specifically, there are four queries focused on impression paintings, four queries on abstract paintings, four queries on still life paintings, and an additional eight queries without any specific limitations. The top-ranked retrieved results by BoonArt and the ArtUK search system for these queries are presented in Figure~\ref{FigureComparisonBoonArtArtUK}. For instance, when queried with `a red cow in the room', BoonArt retrieves the relevant painting image. However, the ArtUK search system only identifies an image with the color red, which is irrelevant to the query. Similarly, when queried with `a man sitting with his dog', BoonArt accurately retrieves a relevant painting image. In contrast, the ArtUK search system only finds an image with a man and his dog, overlooking the critical relation of `sitting with' in the image. 
Figure~\ref{FigureComparisonBoonArtArtUK} highlights the performance of BoonArt in capturing high-level semantics from painting images. 

\subsection{BoonArt's Capability for Image-to-Text Retrieval}

The ArtUK search system does not support the use of image queries, therefore, only the results of BoonArt for image-to-text retrieval are presented. BoonArt utilised eight real-world image queries to retrieve the top-ranked descriptions and their corresponding images from the ArtUK dataset, as shown in Figure~\ref{FigureComparisonBoonArtArtUKi2t}. 
For example, when a query image depicting two trains in a station was provided, BoonArt successfully retrieved the relevant description `two trains in a station' and displayed the corresponding image. Similarly, when a query image of a cat on a chair was given, BoonArt retrieved the relevant description `cat on a chair' along with its corresponding image. These results demonstrate the effectiveness of BoonArt in performing image-to-text retrieval and in facilitating the exploration using visual queries. 

\subsection{Quantitative Results of Cross-modal Information Retrieval with BoonArt} \label{Quantitative Results}

To conduct experiments with quantitative results, the ArtUK dataset was randomly partitioned into three subsets: 5\,783 for training, 500 for validation, and an additional 500 for testing. The evaluation metric used is Recall@$K$ for cross-modal information retrieval experiments, which measures the percentage of relevant items in the top $K$ retrieved results \cite{gong2021limitations, saracevic1995evaluation}. The objective is to retrieve at least one relevant item from a given list, and the average Recall is calculated across all evaluated queries to assess retrieval performance.

Table~\ref{BoonArtUKResults} presents the performance of BoonArt for image-to-text and text-to-image retrieval tasks on the ArtUK test set. Additionally, to highlight the best performance, the results of BoonArt fine-tuned on the training set are also included in Table~\ref{BoonArtUKResults}.  

According to Table~\ref{BoonArtUKResults}, BoonArt (zero-shot) achieved an average Recall@10 value of 93.0\,\% for image-to-text retrieval and 94.2\,\% for text-to-image retrieval in the ArtUK test set. On the other hand, BoonArt (fine-tuned) achieved an average Recall@10 value of 97.0\,\% for image-to-text retrieval and 97.4\,\% for text-to-image retrieval. These findings highlight BoonArt's capability in achieving successful cross-modal information retrieval in the ArtUK dataset.

\begin{table}[htbp]
\caption{Results of cross-modal information retrieval networks on the ArtUK test set. Table shows average Recall@$K$ (\%) values.}
\label{BoonArtUKResults}
\centering
\begin{tabular}{lccccccc}
\hline
\multirow{2}{*}{Engine} & \multirow{2}{*}{Method} & \multicolumn{3}{c}{Image-to-Text [\%]} & \multicolumn{3}{c}{Text-to-Image [\%]} \\
\cmidrule(lr){3-5} \cmidrule(lr){6-8}
                         & & R@1           & R@5           & R@10          & R@1           & R@5           & R@10          \\ \hline
BoonArt           & zero-shot &  68.8  & 88.0 & 93.0 & 71.2 & 90.2 & 94.2  \\

BoonArt            &  fine-tuned & 77.4 & 93.4 & 97.0 & 80.8 & 94.6 & 97.4  \\
\hline
\end{tabular}
\end{table}

\subsection{Evaluation of Retrieval Time}
The average retrieval time for each query by BoonArt was experimentally measured. For the image-to-text retrieval task, which involved processing 6\,783 textual descriptions, the average retrieval time for one query was 0.18 seconds. Similarly, for the text-to-image retrieval task, which included 6\,783 images, the average retrieval time for one query was 0.40 seconds. 

\section{Conclusion}
This paper presents BoonArt, a cross-modal search engine specifically designed for retrieving painting images. To enhance the user experience, BoonArt excels at image-to-text and text-to-image retrieval by extracting high-level visual semantics. It integrates the state-of-the-art VSE network VITR, which focuses on relation-focused cross-modal information retrieval. Extensive experiments were conducted to evaluate BoonArt's performance, demonstrating its ability to outperform the ArtUK search system in text-to-image retrieval tasks. Furthermore, BoonArt surpasses the limitations of the ArtUK search system by enabling image queries, allowing users to retrieve textual descriptions and their corresponding images. BoonArt enhances the retrieval performance of painting images by bridging the gap between textual and visual modalities, resulting in an improved user experience. Currently, the database includes the ArtUK dataset, but it can be extended to work with new datasets. In future work, the database will be expanded to incorporate various datasets of artworks, and the engine will be evaluated using these datasets.

\bibliographystyle{unsrt}  
\bibliography{egbib}  

\begin{thebibliography}{10}

\bibitem{castellano2021visual}
Giovanna Castellano, Eufemia Lella, and Gennaro Vessio.
\newblock Visual link retrieval and knowledge discovery in painting datasets.
\newblock {\em Multimedia Tools and Applications}, 80:6599--6616, 2021.

\bibitem{li2022intuitively}
Kangying Li, Jiayun Wang, Biligsaikhan Batjargal, and Akira Maeda.
\newblock Intuitively searching for the rare colors from digital artwork
  collections by text description: a case demonstration of japanese ukiyo-e
  print retrieval.
\newblock {\em Future Internet}, 14(7):212, 2022.

\bibitem{crowley2014state}
Elliot Crowley and Andrew Zisserman.
\newblock The state of the art: object retrieval in paintings using
  discriminative regions.
\newblock In {\em Proceedings of the British Machine Vision Conference},
  page~8, 2014.

\bibitem{crowley2016art}
Elliot~J Crowley and Andrew Zisserman.
\newblock The art of detection.
\newblock In {\em Proceedings of the European Conference on Computer Vision
  Workshops}, pages 721--737. Springer, 2016.

\bibitem{seidenari2017deep}
Lorenzo Seidenari, Claudio Baecchi, Tiberio Uricchio, Andrea Ferracani, Marco
  Bertini, and Alberto~Del Bimbo.
\newblock Deep artwork detection and retrieval for automatic context-aware
  audio guides.
\newblock {\em ACM Transactions on Multimedia Computing, Communications, and
  Applications}, 13(3s):1--21, 2017.

\bibitem{lowe1999object}
David~G Lowe.
\newblock Object recognition from local scale-invariant features.
\newblock In {\em Proceedings of International Conference on Computer Vision},
  volume~2, pages 1150--1157. Ieee, 1999.

\bibitem{zhang2013image}
Lei Zhang and Yong Rui.
\newblock Image search—from thousands to billions in 20 years.
\newblock {\em ACM Transactions on Multimedia Computing, Communications, and
  Applications}, 9(1s):1--20, 2013.

\bibitem{li2011text}
Wen Li, Lixin Duan, Dong Xu, and Ivor Wai-Hung Tsang.
\newblock Text-based image retrieval using progressive multi-instance learning.
\newblock In {\em International Conference on Computer Vision}, pages
  2049--2055. IEEE, 2011.

\bibitem{kaur2021comparative}
Parminder Kaur, Husanbir~Singh Pannu, and Avleen~Kaur Malhi.
\newblock Comparative analysis on cross-modal information retrieval: a review.
\newblock {\em Computer Science Review}, 39:100336, 2021.

\bibitem{zhang2020context}
Qi~Zhang, Zhen Lei, Zhaoxiang Zhang, and Stan~Z Li.
\newblock Context-aware attention network for image-text retrieval.
\newblock In {\em Proceedings of the IEEE/CVF Conference on Computer Vision and
  Pattern Recognition}, pages 3536--3545, 2020.

\bibitem{li2020unicoder}
Gen Li, Nan Duan, Yuejian Fang, Ming Gong, and Daxin Jiang.
\newblock {Unicoder-VL}: a universal encoder for vision and language by
  cross-modal pre-training.
\newblock In {\em Proceedings of the AAAI Conference on Artificial
  Intelligence}, volume~34, pages 11336--11344, 2020.

\bibitem{young2014image}
Peter Young, Alice Lai, Micah Hodosh, and Julia Hockenmaier.
\newblock From image descriptions to visual denotations: new similarity metrics
  for semantic inference over event descriptions.
\newblock {\em Transactions of the Association for Computational Linguistics},
  2:67--78, 2014.

\bibitem{lin2014microsoft}
Tsung-Yi Lin, Michael Maire, Serge Belongie, James Hays, Pietro Perona, Deva
  Ramanan, Piotr Doll{\'a}r, and C~Lawrence Zitnick.
\newblock Microsoft {COCO}: common objects in context.
\newblock In {\em Proceedings of the European Conference on Computer Vision},
  pages 740--755, 2014.

\bibitem{mao2016generation}
Junhua Mao, Jonathan Huang, Alexander Toshev, Oana Camburu, Alan~L Yuille, and
  Kevin Murphy.
\newblock Generation and comprehension of unambiguous object descriptions.
\newblock In {\em Proceedings of the IEEE/CVF Conference on Computer Vision and
  Pattern Recognition}, pages 11--20, 2016.

\bibitem{faghri2018vse++}
Fartash Faghri, David~J Fleet, Jamie~Ryan Kiros, and Sanja Fidler.
\newblock {VSE++}: improving visual-semantic embeddings with hard negatives.
\newblock In {\em Proceedings of the British Machine Vision Conference},
  page~12, 2018.

\bibitem{anderson2018bottom}
Peter Anderson, Xiaodong He, Chris Buehler, Damien Teney, Mark Johnson, Stephen
  Gould, and Lei Zhang.
\newblock Bottom-up and top-down attention for image captioning and visual
  question answering.
\newblock In {\em Proceedings of the IEEE/CVF Conference on Computer Vision and
  Pattern Recognition}, pages 6077--6086, 2018.

\bibitem{cho2014learning}
Kyunghyun Cho, Bart Van~Merri{\"e}nboer, Caglar Gulcehre, Dzmitry Bahdanau,
  Fethi Bougares, Holger Schwenk, and Yoshua Bengio.
\newblock Learning phrase representations using {RNN} encoder-decoder for
  statistical machine translation.
\newblock In {\em Proceedings of the Conference on Empirical Methods in Natural
  Language Processing}, pages 1724--1734, 2014.

\bibitem{lee2018stacked}
Kuang-Huei Lee, Xi~Chen, Gang Hua, Houdong Hu, and Xiaodong He.
\newblock Stacked cross attention for image-text matching.
\newblock In {\em Proceedings of the European Conference on Computer Vision},
  pages 201--216, 2018.

\bibitem{li2019visual}
Kunpeng Li, Yulun Zhang, Kai Li, Yuanyuan Li, and Yun Fu.
\newblock Visual semantic reasoning for image-text matching.
\newblock In {\em Proceedings of International Conference on Computer Vision},
  pages 4654--4662, 2019.

\bibitem{kipf2017semi}
Thomas~N Kipf and Max Welling.
\newblock Semi-supervised classification with graph convolutional networks.
\newblock In {\em Proceedings of International Conference on Learning
  Representations}, 2017.

\bibitem{chen2021learning}
Jiacheng Chen, Hexiang Hu, Hao Wu, Yuning Jiang, and Changhu Wang.
\newblock Learning the best pooling strategy for visual semantic embedding.
\newblock In {\em Proceedings of the IEEE/CVF Conference on Computer Vision and
  Pattern Recognition}, pages 15789--15798, 2021.

\bibitem{chen2020uniter}
Yen-Chun Chen, Linjie Li, Licheng Yu, Ahmed El~Kholy, Faisal Ahmed, Zhe Gan,
  Yu~Cheng, and Jingjing Liu.
\newblock {UNITER}: universal image-text representation learning.
\newblock In {\em Proceedings of the European Conference on Computer Vision},
  pages 104--120, 2020.

\bibitem{devlin2019bert}
Jacob Devlin, Ming-Wei Chang, Kenton Lee, and Kristina Toutanova.
\newblock {BERT}: pre-training of deep bidirectional transformers for language
  understanding.
\newblock In {\em Proceedings of the Annual Conference of the North American
  Chapter of the Association for Computational Linguistics}, pages 4171--4186,
  2019.

\bibitem{krishna2017visual}
Ranjay Krishna, Yuke Zhu, Oliver Groth, Justin Johnson, Kenji Hata, Joshua
  Kravitz, Stephanie Chen, Yannis Kalantidis, Li-Jia Li, David~A Shamma, et~al.
\newblock Visual genome: connecting language and vision using crowdsourced
  dense image annotations.
\newblock {\em International Journal of Computer Vision}, 123(1):32--73, 2017.

\bibitem{sharma2018conceptual}
Piyush Sharma, Nan Ding, Sebastian Goodman, and Radu Soricut.
\newblock Conceptual captions: a cleaned, hypernymed, image alt-text dataset
  for automatic image captioning.
\newblock In {\em Proceedings of the Annual Meeting of the Association for
  Computational Linguistics}, volume~1, pages 2556--2565, 2018.

\bibitem{ordonez2011im2text}
Vicente Ordonez, Girish Kulkarni, and Tamara Berg.
\newblock Im2text: describing images using 1 million captioned photographs.
\newblock {\em Advances in Neural Information Processing Systems},
  24:1143--1151, 2011.

\bibitem{radford2021learning}
Alec Radford, Jong~Wook Kim, Chris Hallacy, Aditya Ramesh, Gabriel Goh,
  Sandhini Agarwal, Girish Sastry, Amanda Askell, Pamela Mishkin, Jack Clark,
  et~al.
\newblock Learning transferable visual models from natural language
  supervision.
\newblock In {\em Proceedings of International Conference on Machine Learning},
  pages 8748--8763, 2021.

\bibitem{gong2023vitr}
Yan Gong and Georgina Cosma.
\newblock {VITR}: augmenting vision transformers with relation-focused learning
  for cross-modal information retrieval.
\newblock {\em arXiv preprint arXiv:2302.06350}, 2023.

\bibitem{elliot2021artuk}
Ernesto~Coto Elliot J.~Crowley and Andrew Zisserman.
\newblock The {ArtUK} paintings dataset.
\newblock \url{https://www.robots.ox.ac.uk/~vgg/data/paintings/}, 2021.

\bibitem{gong2021limitations}
Yan Gong, Georgina Cosma, and Hui Fang.
\newblock On the limitations of visual-semantic embedding networks for
  image-to-text information retrieval.
\newblock {\em Journal of Imaging}, 7(8):125, 2021.

\bibitem{saracevic1995evaluation}
Tefko Saracevic.
\newblock Evaluation of evaluation in information retrieval.
\newblock In {\em Proceedings of the Annual International ACM SIGIR Conference
  on Research and Development in Information Retrieval}, pages 138--146, 1995.

\end{thebibliography}

\end{document}